\documentclass[runningheads]{llncs}

\usepackage{graphicx}

\usepackage[colorlinks,urlcolor=blue]{hyperref}

\usepackage{marvosym} 
\usepackage{amsfonts} 
\usepackage{booktabs} 
\usepackage{mathtools}
\usepackage{amssymb} 
\usepackage{color} 
\definecolor{higherror}{rgb}{0.3415, 0.062325, 0.429425}
\definecolor{lowerror}{rgb}{0.978422, 0.557937, 0.034931}
\definecolor{pointcolor}{cmyk}{0,0,0,0.5}

\makeatletter
\newcommand{\printfnsymbol}[1]{%
  \textsuperscript{\@fnsymbol{#1}}%
}
\makeatother

\begin{document}
\title{Learning Bloch Simulations for MR Fingerprinting by Invertible Neural Networks}

\titlerunning{Learning Bloch Simulations for MRF by Invertible NNs}

\author{Fabian Balsiger\inst{1,2}\thanks{These authors contributed equally and are listed by flipping a coin.}\orcidID{0000-0001-7577-9870} \and
Alain Jungo\inst{1,2}\printfnsymbol{1}\orcidID{0000-0001-8327-4653} \and
Olivier Scheidegger\inst{3} \and
Benjamin Marty\inst{4,5} \and
Mauricio Reyes\inst{1,2}}
\authorrunning{F. Balsiger and A. Jungo et al.}

\institute{ARTORG Center for Biomedical Research, University of Bern, Bern, Switzerland \\ \email{[fabian.balsiger|alain.jungo]@artorg.unibe.ch}  \and
Insel Data Science Center, Inselspital, Bern University Hospital, Bern, Switzerland \and
Support Center for Advanced Neuroimaging (SCAN), Institute for Diagnostic and Interventional Neuroradiology, Inselspital, Bern University Hospital, Bern, Switzerland \and
NMR Laboratory, Institute of Myology, Neuromuscular Investigation Center, Paris, France \and 
NMR Laboratory, CEA, DRF, IBFJ, MIRCen, Paris, France
}

\maketitle

\begin{abstract}
Magnetic resonance fingerprinting (MRF) enables fast and multiparametric MR imaging. Despite fast acquisition, the state-of-the-art reconstruction of MRF based on dictionary matching is slow and lacks scalability. To overcome these limitations, neural network (NN) approaches estimating MR parameters from fingerprints have been proposed recently. Here, we revisit NN-based MRF reconstruction to jointly learn the forward process from MR parameters to fingerprints and the backward process from fingerprints to MR parameters by leveraging invertible neural networks (INNs). As a proof-of-concept, we perform various experiments showing the benefit of learning the forward process, i.e., the Bloch simulations, for improved MR parameter estimation. The benefit especially accentuates when MR parameter estimation is difficult due to MR physical restrictions. Therefore, INNs might be a feasible alternative to the current solely backward-based NNs for MRF reconstruction.


\keywords{Reconstruction \and Magnetic resonance fingerprinting \and Invertible neural network.}
\end{abstract}

\setcounter{footnote}{0}

\section{Introduction}

Magnetic resonance fingerprinting (MRF)~\cite{Ma2013} is a relatively new but increasingly used~\cite{Poorman2019} concept for fast and multiparametric quantitative MR imaging. Acquisitions of MRF produce unique magnetization evolutions per voxel, called fingerprints, due to temporal varying MR sequence schedules. From these fingerprints, MR parameters (e.g., relaxation times) are then reconstructed using a dictionary matching, comparing each fingerprint to a dictionary of simulated fingerprints with known MR parameters. Although the MRF acquisition itself is fast thanks to high undersampling, the dictionary matching is slow, discrete and cannot interpolate, and lacks scalability with increasing number of MR parameters.

With the advent of deep learning, neural networks (NNs) have been explored to overcome the limitations of the dictionary matching. The dictionary matching can be formulated as a regression problem from the fingerprints to the MR parameters. Several methods have been applied to MRF with impressive results both in terms of reconstruction accuracy and speed \cite{Cohen2018,Hoppe2017,Fang2019,Oksuz2019,Balsiger2019a,Golbabaee2019,Song2019,Hoppe2019,Fang2019b,Balsiger2020a}. Among these, spatially regularizing methods trained on \textit{in vivo} MRF acquisitions showed superiority over methods performing fingerprint-wise regression \cite{Balsiger2018b,Fang2019,Balsiger2019a,Balsiger2020a,Fang2019b,Hoppe2019}. However, spatial methods might require a considerable amount of training data to achieve reasonable robustness for highly heterogeneous diseases~\cite{Balsiger2020a}. Therefore, robust fingerprint-wise methods, leveraging the dictionaries for training, are required to alleviate the need of \textit{in vivo} MRF acquisitions.

We revisit NN-based MRF reconstruction by formulating it as an inverse problem where we jointly learn the forward process from MR parameters to fingerprints and the backward process from fingerprints to MR parameters. In doing so, the available information of the forward process is leveraged, which might help disentangling MR physical processes and consequently improve the MR parameter estimation of the backward process. To this end, we leverage invertible neural networks (INNs)~\cite{Dinh2017}. As proof-of-concept, we perform various experiments showing the benefit of learning the forward process, i.e., the Bloch simulations, for improved NN-based MRF reconstruction.

\section{Methodology}

\subsection{MR Fingerprinting using Invertible Neural Networks}
Inverse problems are characterized by having some observations $\mathbf{y}$, from which we want to obtain the underlying parameters $\mathbf{x}$. The forward process $\mathbf{y} = f(\mathbf{x})$ is usually well defined and computable. However, the backward process $\mathbf{x} = f^{-1}(\mathbf{y})$ is not trivial to compute. MRF can be formulated as an inverse problem~\cite{Boux2019}. The forward process $f$ is described by the Bloch equations~\cite{Bloch1946}. Meaning, from some MR parameters $\mathbf{x} \in \mathbb{R}^M$, one can simulate a corresponding fingerprint $\mathbf{y} \in \mathbb{C}^T$ for a given MRF sequence. The backward process $f^{-1}$ in MRF is typically solved by dictionary matching, or recently via regression by NNs. However, in doing so, the knowledge of the well-defined forward process is completely omitted in the backward process. We hypothesize that by leveraging the knowledge of the forward process, NN-based MRF reconstruction can be improved. Therefore, we aim at jointly learning the forward and the backward process by using INNs. Once learned, the trained INN can be used to estimate MR parameters $\mathbf{x}$ from a fingerprint $\mathbf{y}$, as done in literature.

\begin{figure}[t]
\centering
\includegraphics[width=.9\textwidth]{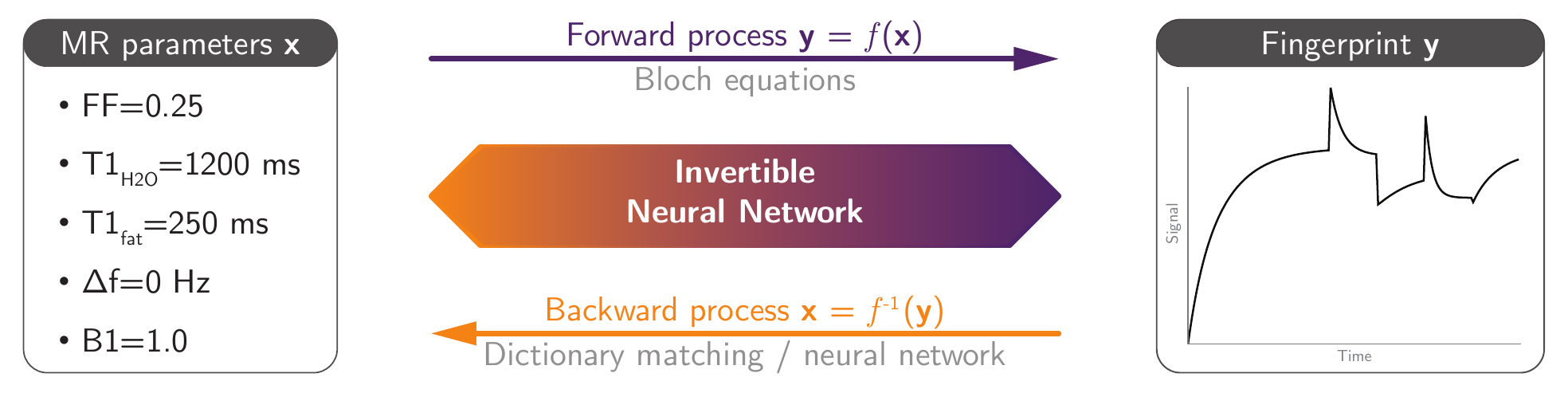}
\caption{Overview of the INN in the context of MRF. The forward process simulates fingerprints $\mathbf{y}$ from MR parameters $\mathbf{x}$, usually by Bloch simulations. The backward process estimates MR parameters $\mathbf{x}$ of a fingerprint $\mathbf{y}$, usually by dictionary matching or recently NNs. The INN is capable of doing both the forward and backward process.}
\label{fig-overview}
\end{figure}

Fig.~\ref{fig-overview} depicts INNs in the context of MRF. Given training pairs $(\mathbf{x}, \mathbf{y})$ from a dictionary, the MR parameters $\mathbf{x}$ are fed into the INN, which predicts the fingerprint $\mathbf{\hat{y}}$. Optimizing a mean squared error (MSE) loss between $\mathbf{y}$ and $\mathbf{\hat{y}}$ results in learning the forward process. Feeding the fingerprint $\mathbf{y}$ from the opposite direction into the INN predicts the MR parameters $\mathbf{\hat{x}}$. Here, we also optimize a MSE loss between $\mathbf{x}$ and $\mathbf{\hat{x}}$ to learn the backward process\footnote{The MSE loss was empirically found to be beneficial although the backward process is theoretically learned through the bijectivity property of the INN.}. For both forward and backward, the INN uses the same weights, and, therefore, the training jointly optimizes the forward and backward process.

The architecture of our INN bases on RealNVP~\cite{Dinh2017} and consists of two reversible blocks with permutation layers~\cite{Ardizzone2019}. A reversible block is composed of two complementary affine transformations, with scales $s_i$ and translations $t_i$ ($i \in \{ 1, 2 \}$). The transformations describe the forward pass as

\begin{equation}
\mathbf{v}_{1} = \mathbf{u}_{1} \odot \exp (s_{2} (\mathbf{u}_{2} ))+t_{2}(\mathbf{u}_{2})\,, \quad \mathbf{v}_{2}=\mathbf{u}_{2} \odot \exp (s_{1}(\mathbf{v}_{1}))+t_{1}(\mathbf{v}_{1})\,,
\nonumber
\end{equation}

\noindent where $\mathbf{u} = [\mathbf{u}_{1}$, $\mathbf{u}_{2}]$ and $\mathbf{v} = [\mathbf{v}_{1}$, $\mathbf{v}_{2}]$ are the input and output split into halves, and $\odot$ is the Hadamard product. The reversibility of the affine transformations ensure the invertibility of the reversible block, such that the inverse is given by
\begin{equation}
\mathbf{u}_{2} = (\mathbf{v}_{2}-t_{1}(\mathbf{v}_{1})) \odot \exp \left(-s_{1}\left(\mathbf{v}_{1}\right)\right)\,, \quad \mathbf{u}_{1}=(\mathbf{v}_{1}-t_{2}(\mathbf{u}_{2})) \odot \exp (-s_{2}(\mathbf{u}_{2}))\,.
\nonumber
\end{equation}

\noindent As a consequence, the operations $s$ and $t$ do not need to be invertible themselves. For each $s_i$ and $t_i$, we use two fully-connected layers, with 128 neurons each, followed by ReLU and linear activation, respectively. The permutation layers enforce a different split of the halves in every reversible block~\cite{Ardizzone2019}. We remark that we zero-pad the input $\mathbf{x}$ to match the dimensionality of $\mathbf{y}$. 
Generally, such INN architectures have been shown to be suitable to solve diverse inverse problems~\cite{Ardizzone2019}, including problems in medical imaging~\cite{Adler2019}.

\subsection{MR Fingerprinting Sequence}
In the context of our clinical scope, we use MRF T1-FF~\cite{Marty2019a}, a MRF sequence designed for the quantification of T1 relaxation time (T1) and fat fraction (FF) in fatty infiltrated tissues such as diseased skeletal muscle. Since fat can heavily bias the T1 quantification, MRF T1-FF separately estimates the T1 of water (T1\textsubscript{H2O}) and T1 of fat (T1\textsubscript{fat}) pools. Additionally, the confunding effects of static magnetic field inhomogeneity ($\Delta$f) and flip angle efficacy (B1) are quantified, resulting in a total of $M=5$ MR parameters (FF, T1\textsubscript{H2O}, T1\textsubscript{fat}, $\Delta$f, and B1). Fingerprints are simulated using the Bloch equations with varying MRF sequence schedules of flip angles, echo times and repetition times, resulting in fingerprints of length $T=175$.

\label{sec:dictionary}
Two dictionaries were simulated, one for training and the other for validation and testing. The training dictionary was simulated with (start:increment:stop) $(0.0{:}0.1{:}1.0)$ for FF, $(500{:}100{:}1700,1900{:}200{:}3100)$~ms for T1\textsubscript{H2O}, $(200{:}25{:}400)$~ms for T1\textsubscript{fat}, $(-120{:}10{:}120)$~Hz for $\Delta$f, and $(0.3{:}0.1{:}1.0)$ for B1. The other dictionary was simulated with $(0.05{:}0.1{:}0.95)$ for FF, $(550{:}200{:}1750,2150{:}400{:}2950)$~ms for T1\textsubscript{H2O}, $(215{:}50{:}365)$~ms for T1\textsubscript{fat}, $(-115{:}20{:}105)$~Hz for $\Delta$f, and $(0.35{:}0.1{:}0.95)$ for B1, of which randomly 20~\% of the entries were used for validation and the remaining 80~\% for testing. In total, 396000 entries were used for training, 6720 for validation, and 26880 unseen entries for testing.

\subsection{Baselines and Training}
We compared the INN to five baselines, one ablation and four competing NN-based methods. The ablation, termed INN\textsubscript{bwd}, uses exactly the same architecture as INN but was only trained on the backward process to ablate the benefit of jointly learning the forward and backward process. The competing methods are: (i) a fully-connected NN by Cohen et al.~\cite{Cohen2018} with two hidden layers, (ii) a NN by Hoppe et al.~\cite{Hoppe2019} consisting of four convolution layers followed by four fully-connected layer, (ii) a recurrent NN by Oksuz et al.~\cite{Oksuz2019} based on gated recurrent units with 100 recurrent layers followed by a fully-connected layer, and (iv) a 1-D residual convolutional NN by Song et al.~\cite{Song2019}.

All NNs were trained using a MSE loss with an Adam optimizer~\cite{Kingma2015} with the learning rate chosen from $\{0.01, 0.001, 0.0005, 0.0001 \}$, and $\beta_1 = 0.9, \beta_2 = 0.999$. We trained for 80 epochs and chose the batch size from $\{50, 200\}$. At each epoch, the coefficient of determination (R\textsuperscript{2}) between $\mathbf{x}$ and $\mathbf{\hat{x}}$ on the validation set was calculated and the best model was used for testing. As input, the real and imaginary parts of the complex-valued fingerprints $\mathbf{y}$ were concatenated, as commonly done~\cite{Balsiger2018b,Fang2019,Balsiger2019a,Balsiger2020a,Hoppe2019}, resulting in an input dimension of $2T = 350$ in all experiments. The output dimension was $M = 5$, resulting in a zero padding of $\mathbf{x}$ for the INN of $2T-M = 345$. As data augmentation, the fingerprints $\mathbf{y}$ were perturbed with random noise $\mathcal{N}(0, N^2)$. The noise standard deviation $N$ was set to imitate signal-to-noise ratio (SNR) conditions of MRF T1-FF scans. The SNR (in dB) was defined as $20 \log_{10}( S / N)$, where $S$ is the mean intensity of the magnitude of the magnetization at thermal equilibrium in healthy skeletal muscle. $N$ was set to $0.003$ for training, and $\mathbf{y}$ was perturbed for both the forward and backward process when training the INN. As no public code was available for the competing NNs, we implemented them in PyTorch 1.3 along with the INN. We release the code at \url{http://www.github.com/fabianbalsiger/mrf-reconstruction-mlmir2020}.

\section{Experiments and Results}

\subsection{Backward Process: MR Parameter Estimation}
The results of the MR parameter estimation from unperturbed fingerprints $\mathbf{y}$ are summarized in Table~\ref{tab-quantiative}. The mean absolute error (MAE), the mean relative error (MRE), and the R\textsuperscript{2} between the reference $\mathbf{x}$ and predicted $\mathbf{\hat{x}}$ MR parameters were calculated. The INN estimated all MR parameters with the highest accuracy except for the MR parameter $\Delta$f, where the INN\textsubscript{bwd} yielded the best estimations in terms of MAE. Overall, all methods performed in a similar range for FF, $\Delta$f, and B1. However, a benefit in learning the Bloch simulations accentuated especially for T1\textsubscript{H2O} and T1\textsubscript{fat}, where the INN outperformed all competing methods including the ablation by a considerable margin. We analyze this behaviour in more detail in Sec.~\ref{sec:fwd-process}.

\begin{table}[!b]
	\tiny
	\caption{Mean absolute error (MAE), mean relative error (MRE), and the coefficient of determination (R\textsuperscript{2}) of the MR parameter estimation from unperturbed fingerprints. a.u.: arbitrary unit.}
	\label{tab-quantiative}
	\centering
	\scalebox{1.07}{
	\begin{tabular}{ll@{\hspace{0.5em}}c@{\hspace{1em}}c@{\hspace{1em}}c@{\hspace{1em}}c@{\hspace{1em}}c@{\hspace{1em}}c}
		\toprule
		& & \multicolumn{6}{c}{Method} \\
		\cmidrule{3-8}
		Metric & MR parameter & INN & INN\textsubscript{bwd} & Cohen et al. & Hoppe et al. & Oksuz et al. & Song et al.\\
		\midrule
		MAE & FF & \textbf{0.008}\scalebox{0.75}{$\,\pm\,$0.007} & 0.013\scalebox{0.75}{$\,\pm\,$0.010} & 0.013\scalebox{0.75}{$\,\pm\,$0.011} & 0.016\scalebox{0.75}{$\,\pm\,$0.012} & 0.015\scalebox{0.75}{$\,\pm\,$0.012} & 0.015\scalebox{0.75}{$\,\pm\,$0.012} \\
		 & T1\textsubscript{H2O}\scalebox{0.75}{ (ms)} & \textbf{88.9}\scalebox{0.75}{$\,\pm\,$170.2} & 143.2\scalebox{0.75}{$\,\pm\,$249.3} & 140.6\scalebox{0.75}{$\,\pm\,$234.8} & 162.2\scalebox{0.75}{$\,\pm\,$241.8} & 176.0\scalebox{0.75}{$\,\pm\,$239.8} & 160.1\scalebox{0.75}{$\,\pm\,$243.5} \\
		 & T1\textsubscript{fat}\scalebox{0.75}{ (ms)} & \textbf{20.8}\scalebox{0.75}{$\,\pm\,$19.4} & 27.8\scalebox{0.75}{$\,\pm\,$21.6} & 27.9\scalebox{0.75}{$\,\pm\,$22.0} & 29.0\scalebox{0.75}{$\,\pm\,$22.1} & 31.7\scalebox{0.75}{$\,\pm\,$23.0} & 28.1\scalebox{0.75}{$\,\pm\,$22.8} \\
		 & $\Delta$f\scalebox{0.75}{ (Hz)} & 0.736\scalebox{0.75}{$\,\pm\,$0.666} & \textbf{0.665}\scalebox{0.75}{$\,\pm\,$0.490} & 0.833\scalebox{0.75}{$\,\pm\,$0.612} & 2.635\scalebox{0.75}{$\,\pm\,$1.503} & 1.380\scalebox{0.75}{$\,\pm\,$1.083} & 1.532\scalebox{0.75}{$\,\pm\,$1.169} \\
		 & B1\scalebox{0.75}{ (a.u.)} & \textbf{0.012}\scalebox{0.75}{$\,\pm\,$0.010} & 0.013\scalebox{0.75}{$\,\pm\,$0.010} & 0.015\scalebox{0.75}{$\,\pm\,$0.013} & 0.016\scalebox{0.75}{$\,\pm\,$0.014} & 0.027\scalebox{0.75}{$\,\pm\,$0.021} & 0.019\scalebox{0.75}{$\,\pm\,$0.014} \\
		\midrule
		MRE & FF\scalebox{0.75}{ (\%)} & \textbf{2.89}\scalebox{0.75}{$\,\pm\,$4.69} & 4.23\scalebox{0.75}{$\,\pm\,$5.62} & 4.09\scalebox{0.75}{$\,\pm\,$5.02} & 5.64\scalebox{0.75}{$\,\pm\,$7.62} & 5.10\scalebox{0.75}{$\,\pm\,$6.99} & 6.33\scalebox{0.75}{$\,\pm\,$11.94} \\
		 & T1\textsubscript{H2O}\scalebox{0.75}{ (\%)} & \textbf{6.75}\scalebox{0.75}{$\,\pm\,$15.46} & 11.55\scalebox{0.75}{$\,\pm\,$27.23} & 11.47\scalebox{0.75}{$\,\pm\,$25.21} & 12.66\scalebox{0.75}{$\,\pm\,$25.22} & 13.32\scalebox{0.75}{$\,\pm\,$23.07} & 13.36\scalebox{0.75}{$\,\pm\,$27.42} \\
		 & T1\textsubscript{fat}\scalebox{0.75}{ (\%)} & \textbf{7.48}\scalebox{0.75}{$\,\pm\,$7.28} & 10.34\scalebox{0.75}{$\,\pm\,$9.22} & 10.33\scalebox{0.75}{$\,\pm\,$9.40} & 10.97\scalebox{0.75}{$\,\pm\,$9.80} & 11.96\scalebox{0.75}{$\,\pm\,$10.27} & 10.41\scalebox{0.75}{$\,\pm\,$9.81} \\
		 & $\Delta$f\scalebox{0.75}{ (\%)} & \textbf{2.50}\scalebox{0.75}{$\,\pm\,$5.00} & 2.86\scalebox{0.75}{$\,\pm\,$6.07} & 3.16\scalebox{0.75}{$\,\pm\,$5.91} & 7.52\scalebox{0.75}{$\,\pm\,$11.40} & 3.71\scalebox{0.75}{$\,\pm\,$5.40} & 5.19\scalebox{0.75}{$\,\pm\,$9.00} \\
		 & B1\scalebox{0.75}{ (\%)} & \textbf{1.98}\scalebox{0.75}{$\,\pm\,$1.95} & 2.17\scalebox{0.75}{$\,\pm\,$1.88} & 2.56\scalebox{0.75}{$\,\pm\,$2.34} & 2.83\scalebox{0.75}{$\,\pm\,$2.91} & 4.22\scalebox{0.75}{$\,\pm\,$3.01} & 3.18\scalebox{0.75}{$\,\pm\,$2.69} \\
		\midrule
		R\textsuperscript{2} & FF & \textbf{0.999} & 0.997 & 0.996 & 0.995 & 0.995 & 0.995 \\
		 & T1\textsubscript{H2O} & \textbf{0.934} & 0.852 & 0.866 & 0.848 & 0.841 & 0.848 \\
		 & T1\textsubscript{fat} & \textbf{0.741} & 0.604 & 0.596 & 0.574 & 0.508 & 0.582 \\
		 & $\Delta$f & \textbf{1.000} & \textbf{1.000} & \textbf{1.000} & 0.998 & 0.999 & 0.999 \\
		 & B1 & \textbf{0.994} & 0.993 & 0.990 & 0.988 & 0.972 & 0.986 \\
		\bottomrule
	\end{tabular}
	}
\end{table}

\begin{figure}[t]
\centering
\includegraphics[width=\textwidth]{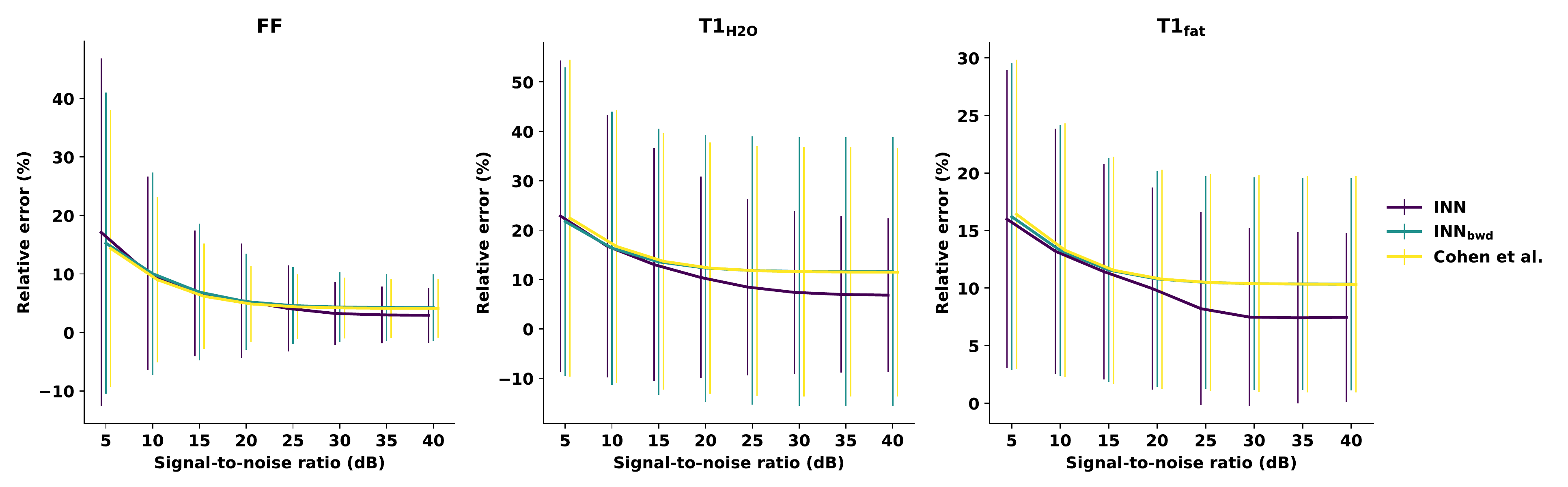}
\caption{Reconstruction performance in mean relative error of the INN, INN\textsubscript{bwd}, and Cohen et al.~\cite{Cohen2018} under varying SNR conditions for the MR parameters FF, T1\textsubscript{H2O}, and T1\textsubscript{fat}. Error bars indicate $\pm$ standard deviation. An SNR level of approximately 20 dB can be considered similar to an \textit{in vivo} MRF T1-FF scan.}
\label{fig-snr-reconstruction}
\end{figure}

Robustness to noise is of considerable importance for MRF reconstruction applied to \textit{in vivo} MRF acquisitions due to high undersampling. To simulate undersampling conditions, the performance of the INN, the INN\textsubscript{bwd}, and the best competing method (Cohen et al.~\cite{Cohen2018}) were analyzed under varying SNR levels, see Fig.~\ref{fig-snr-reconstruction}. For each SNR level, we performed Monte Carlo simulations perturbing the fingerprints $\mathbf{y}$ with 100 random noise samples. It is notable that the INN more accurately and precisely estimated the MR parameters at higher SNR levels ($>$ 25~dB) than the other methods. At lower SNR levels, the differences between the methods became negligible, indicating that the benefit of learning the forward pass vanishes as the noise level increases. The plots for the MR parameters $\Delta$f and B1 look similar, and are omitted due to space constraints.

The inference time of the INN was approximately 50~milliseconds for 1000 fingerprints, which is in-line with the competing methods. Only the training time was approximately doubled with 5~minutes for one epoch compared to the competing methods. The number of parameters were 0.36 million for the INN and INN\textsubscript{bwd}, 0.20 million for Cohen et al.~\cite{Cohen2018}, 6.56 million for Hoppe et al.~\cite{Hoppe2019}, 0.12 million for Oksuz et al.~\cite{Oksuz2019}, and 1.49 million for Song et al.~\cite{Song2019}.

\subsection{Forward Process: Benefit of Learning Bloch Simulations}
\label{sec:fwd-process}
\begin{figure}[t]
	\centering
	\includegraphics[width=\textwidth]{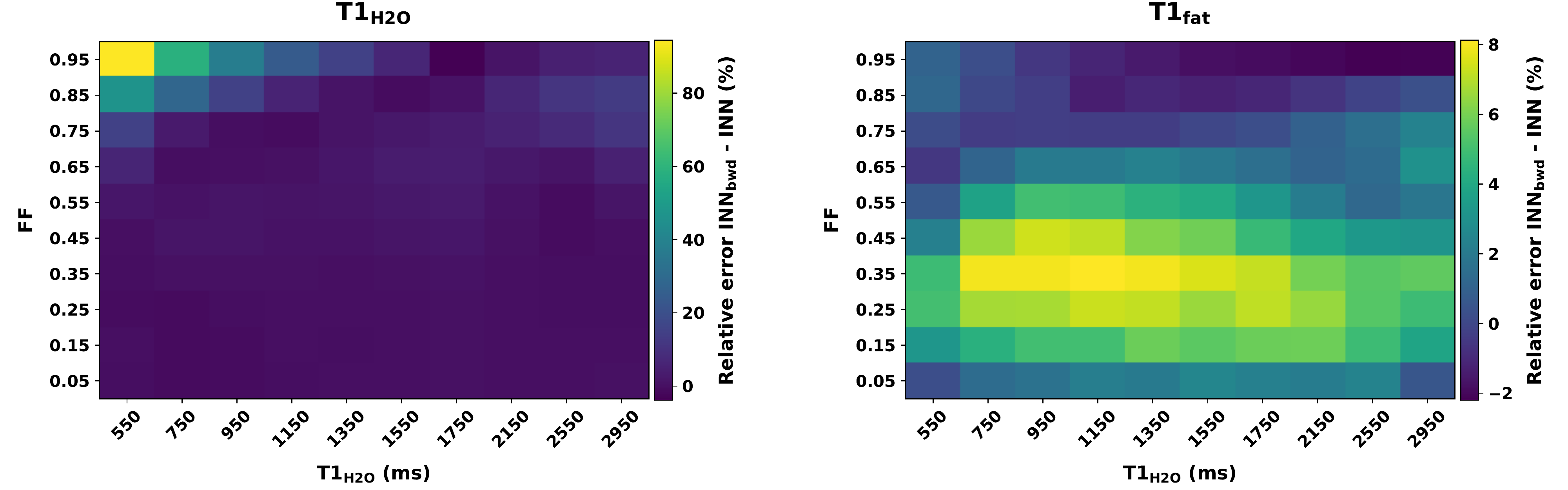}
	\caption{Heat maps of the relative error differences between INN\textsubscript{bwd} and INN for the MR parameters T1\textsubscript{H2O} (left) and T1\textsubscript{fat} (right). Positive values indicate better performance of the INN.}
	\label{fig-error-difference}
\end{figure}

Jointly learning the forward process mainly benefits estimating T1\textsubscript{H2O} and T1\textsubscript{fat} (cf. Table~\ref{tab-quantiative}). To analyze this benefit, we need to introduce MR physics in presence of fat. The used sequence MRF T1-FF is designed for T1 quantification in fatty infiltrated tissues where the fat infiltration occurs at varying fractions, from no fat (FF=0.0), to being solely fat (FF=1.0). Unfortunately, fat infiltration, and therefore FF, greatly affects T1 quantification~\cite{Marty2018a}. At FF=1.0, T1\textsubscript{H2O} is not measurable as no water is present. Similarly, at FF=0.0, T1\textsubscript{fat} is not measurable as no fat is present. Generally, estimating T1\textsubscript{H2O} is difficult at high FF values as the pooled (or global) T1 is heavily biased by the T1\textsubscript{fat}. Contrarily, at low FF values, estimating T1\textsubscript{fat} is difficult as almost no fat is present. Learning the forward process could especially benefit such cases, i.e., when the information in the fingerprints is ambiguous due to MR physical restrictions. To test this assumption, we calculated the difference between the relative errors of the INN\textsubscript{bwd} and INN. The heat maps in Fig.~\ref{fig-error-difference} show the differences for estimated T1\textsubscript{H2O} and T1\textsubscript{fat} at varying FF and T1\textsubscript{H2O} values. On the one hand, the forward process helped at estimating short T1\textsubscript{H2O} ($<$ 1000~ms) at high FF more accurately than INN\textsubscript{bwd}, Fig.~\ref{fig-error-difference} left. Short T1\textsubscript{H2O} values are especially difficult to differentiate from T1\textsubscript{fat}, as these are also very short (cf. dictionary ranges in Sec.~\ref{sec:dictionary}). On the other hand, the forward process benefited the estimation of T1\textsubscript{fat} values at lower FF ($<$ 0.5), Fig.~\ref{fig-error-difference} right. At the very low FF of 0.05, the benefit diminished as it seems difficult to discriminate short T1\textsubscript{fat} values from longer T1\textsubscript{H2O} values, even when the forward process was learned. A nearly identical pattern was also obtained when comparing the INN with the method of Cohen et al.~\cite{Cohen2018} (not shown). These results indicate that learning the forward process helps disentangling underlying MR physical processes.

\subsection{Relation between the Forward and Backward Process}
\begin{figure}
	\centering
	\includegraphics[width=\textwidth]{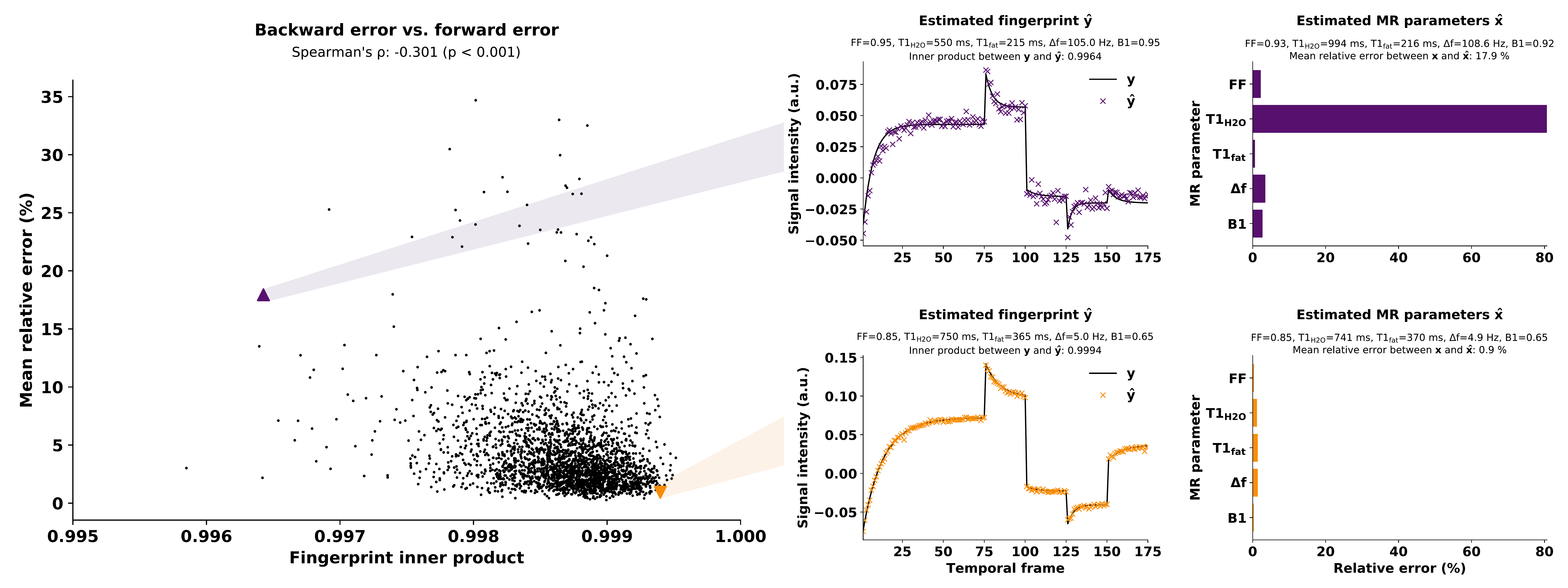}
	\caption{Scatter plot relating the error in the forward and the backward process. For a pair ($\mathbf{x}$, $\mathbf{y}$), we calculated ($\mathbf{\hat{x}}$, $\mathbf{\hat{y}}$) using the INN and plotted the mean relative error between $\mathbf{x}$ and $\mathbf{\hat{x}}$ versus the inner product between $\mathbf{y}$ and $\mathbf{\hat{y}}$. The fingerprints and the relative errors of a high (\textcolor{higherror}{$\blacktriangle$}) and a low (\textcolor{lowerror}{$\blacktriangledown$}) error example are shown on the right-hand side. For visualization purposes, only a random subset of 10~\% of the data points in the scatter plot and the real part of the fingerprints were plotted. a.u.: arbitrary unit.}
	\label{fig-error-correlation}
\end{figure}

Learning the Bloch simulations benefits not only the MR parameter estimations but could also foster interpretability of the estimation. Due to the cyclic nature of the INN, large errors in the backward process, i.e., the error between $\mathbf{x}$ and $\mathbf{\hat{x}}$, should be associated with large errors in the forward process, i.e., the error between $\mathbf{y}$ and $\mathbf{\hat{y}}$. We tested this hypothesis by analyzing the correlation of the MRE between $\mathbf{x}$ and $\mathbf{\hat{x}}$ and the inner product between the fingerprints $\mathbf{y}$ and $\mathbf{\hat{y}}$. The association between the MRE and the inner product is shown in the scatter plot of Fig.~\ref{fig-error-correlation}. The Spearman rank-order correlation coefficient was -0.301 (p $<$ 0.001), indicating a weak monotonic relationship. A high and a low error example are shown on the right-hand side of Fig.~\ref{fig-error-correlation}. The lower agreement between $\mathbf{y}$ and $\mathbf{\hat{y}}$ of the high error example is visually noticeable compared to the low error example. The main source of error is the T1\textsubscript{H2O}, which is difficult to estimate at the high FF of 0.95 the fingerprint $\mathbf{y}$ was simulated with.

\section{Discussion and Conclusion}
We revisited NN-based MRF reconstruction by formulating it as an inverse problem. The INN allows to jointly learn the forward process from MR parameters to fingerprints and the backward process from fingerprints to MR parameters. Regarding reconstruction performance, our results suggest that learning the Bloch simulations is beneficial for MR parameter estimation.

Our experiments showed that the benefit of the INN is considerable when the information in the fingerprints is ambiguous due to MR physical restrictions. Independent of the method (invertible, fully-connected, convolutional, or recurrent) and the network size (number of parameters), FF, $\Delta$f, and B1 were nearly identically well estimated. The errors for these MR parameters were below a step size to simulate dictionaries of reasonable size for the computational intensive dictionary matching. However, this is not the case for T1\textsubscript{H2O} and T1\textsubscript{fat}, where the INN performs superior. By ablation, we could attribute this performance gain to the learning of the forward process. This insight might have implications beyond T1 and FF quantification, e.g., for fast imaging with steady-state precession (FISP) sequences, where T2 relaxation time quantification is more difficult than T1 quantification~\cite{Cohen2018,Fang2019,Hoppe2019,Fang2019b}. Further, the interplay between the forward and backward process enable an enhanced interpretability of the method, which might be regarded as reconstruction uncertainty. This might be useful for MRF sequence design and optimization targeted to NN-based reconstruction.

The main limitation of this proof-of-concept study is clearly that the method was not applied to \textit{in vivo} MRF acquisitions. Prior to doing, the behaviour of the INN under heavy noise conditions needs to be further investigated. It is currently unclear, as to why the benefit of the forward process diminishes at lower SNR levels (cf. Fig.~\ref{fig-snr-reconstruction}). The simplest explanation is clearly the lack of enough signal, which makes MR parameter estimation difficult, independent of the method. Here, spatial regularization would most likely help~\cite{Balsiger2019a,Balsiger2020a}, which is also possible with INNs. First attempts in this direction are promising.

In conclusion, we showed that jointly learning the forward and backward process benefits the reconstruction of MRF. INNs are suitable for such joint learning and might be a feasible alternative to the current solely backward-based NNs for MRF reconstruction.

\subsubsection{\ackname}
This research was supported by the Swiss National Science Foundation (SNSF). The authors thank the NVIDIA Corporation for their GPU donation.

\bibliographystyle{splncs04}
\bibliography{library_short}
\end{document}